\documentclass[useAMS,usenatbib,usegraphicx]{mn2e}
\usepackage[breaklinks,colorlinks,citecolor=black,linkcolor=black,urlcolor=black]{hyperref}

\newcommand*{\figscale}{0.9}

\newcommand*{\mysub}[2]{\ensuremath{#1_{\mathrm{#2}}}}
\newcommand*{\unit}[1]{\ensuremath{\mathrm{\, #1}}}

\newcommand*{\Omegam}{\mysub{\Omega}{m}}

\newcommand*{\Mgas}{\mysub{M}{gas}}

\newcommand*{\Yx}{\mysub{Y}{X}}

\newcommand*{\mproton}{\mysub{m}{p}}
\newcommand*{\rhocr}{\mysub{\rho}{c}}
\newcommand*{\rc}{\mysub{r}{c}}
\newcommand*{\rt}{\mysub{r}{t}}
\newcommand*{\rmax}{\mysub{r}{max}}

\newcommand*{\second}{\unit{s}}

\newcommand*{\km}{\unit{km}}
\newcommand*{\kpc}{\unit{kpc}}
\newcommand*{\Mpc}{\unit{Mpc}}
\newcommand*{\keV}{\unit{keV}}
\newcommand*{\Msun}{\ensuremath{\, M_{\odot}}}

\newcommand*{\expectation}[1]{\ensuremath{\left\langle #1 \right\rangle}}
\newcommand*{\ltsim}{\ {\raise-.75ex\hbox{$\buildrel<\over\sim$}}\ }
\newcommand*{\gtsim}{\ {\raise-.75ex\hbox{$\buildrel>\over\sim$}}\ }
\newcommand*{\proptosim}{\ {\raise-.75ex\hbox{$\buildrel\propto\over\sim$}}\ }

\newcommand*{\secref}[1]{Section~\ref{#1}}

\newcommand*{\appref}[1]{Appendix~\ref{#1}}
\newcommand*{\eqnref}[1]{Equation~\ref{#1}}

\newcommand*{\figref}[1]{Figure~\ref{#1}}
\newcommand*{\tabref}[1]{Table~\ref{#1}}

\newcommand*{\Chandra}{{\it Chandra}}
\newcommand*{\Suzaku}{{\it Suzaku}}

\defcitealias{Navarro9611107}{NFW}
\newcommand*{\NFW}{\citetalias{Navarro9611107}}

\title[Priors on cluster masses and scaling relations]{Implicit Priors in Galaxy Cluster Mass and Scaling Relation Determinations}

\author[A. Mantz and S. W. Allen]{
  A.~Mantz$^{1}$\thanks{E-mail: \href{mailto: amantz@slac.stanford.edu}{amantz@slac.stanford.edu}} and S.~W.~Allen$^{2,3}$\\
  $^1$NASA Goddard Space Flight Center, Code 662, Greenbelt, MD 20771, USA\\
  $^2$Kavli Institute for Particle Astrophysics and Cosmology, Stanford University, 452 Lomita Mall, Stanford, CA 94305-4085, USA\\
  $^3$SLAC National Accelerator Laboratory, 2575 Sand Hill Road, Menlo Park, CA 94025, USA.
}

\date{9 June 2011}

\begin{document}

\pagerange{\pageref{firstpage}--\pageref{lastpage}} \pubyear{2011}
\maketitle
\label{firstpage}

\begin{abstract}
  Deriving the total masses of galaxy clusters from observations of the intracluster medium (ICM) generally requires some prior information, in addition to the assumptions of hydrostatic equilibrium and spherical symmetry. Often, this information takes the form of particular parametrized functions used to describe the cluster gas density and temperature profiles. In this paper, we investigate the implicit priors on hydrostatic masses that result from this fully parametric approach, and the implications of such  priors for scaling relations formed from those masses. We show that the application of such fully parametric models of the ICM naturally imposes a prior on the slopes of the derived scaling relations, favoring the self-similar model, and argue that this prior may be influential in practice. In contrast, this bias does not exist for techniques which adopt an explicit prior on the form of the mass profile but describe the ICM non-parametrically. Constraints on the slope of the cluster mass--temperature relation in the literature show a separation based the approach employed, with the results from fully parametric ICM modeling clustering nearer the self-similar value. Given that a primary goal of scaling relation analyses is to test the self-similar model, the application of methods subject to strong, implicit priors should be avoided. Alternative methods and best practices are discussed.
\end{abstract}

\nokeywords

\section{Introduction} \label{sec:intro}

Scaling relations between observable properties and total gravitating mass are a critical ingredient for cosmological tests based on galaxy clusters (for a review, see \citealt*{Allen1103.4829}). For tests using the abundance, clustering and growth of clusters, scaling relations provide essential mass proxies and are fundamentally important in accounting for selection biases. Our knowledge of these relations and the systematics that affect them currently limits the achievable constraints on some cosmological parameters (e.g. \citealt{Mantz0709.4294,Mantz0909.3099,Mantz0909.3098,Vikhlinin0805.2207,Vikhlinin0812.2720,Rozo0902.3702,Wu0907.2690}, and references therein). Measurements of cluster gas mass fractions, which constrain the mean matter density cosmic expansion history (e.g. \citealt{Sasaki9611033,Allen0405340,Allen0706.0033}), can also be expressed in terms of a scaling relation, namely gas mass as a function of total mass. In addition, the scaling relations are of considerable astrophysical interest, reflecting the complex response of the baryonic components of these systems to their overall gravitational potentials, environments and formation histories. For example, departures from the self-similar form introduced by \citet{Kaiser1986MNRAS.222..323K} provide clues to non-gravitational processes at work in clusters (e.g. \citealt{Voit0410173} and references therein).

In previous work, we have emphasized the need to model covariance between measured quantities in the  analysis of cluster scaling relations \citep{Mantz0909.3099,Mantz0909.3098}. Here, we distinguish  further between various contributing factors to such covariance:
\begin{enumerate}
\item Covariance that is intrinsic to the measurement process. For example, when multiple quantities are measured from an X-ray observation, Poisson uncertainties due to photon counting affect these quantities in a coherent rather than independent way.
\item Covariance due to explicit use of one measurement to inform another. For example, if the Compton $Y$ signal from a Sunyaev-Zel'dovich observation is measured within a radius determined from an X-ray observation, the statistical error in the radius determination coherently affects the errors on $Y$ and on any X-ray quantities measured within that radius.
\item Covariance that is introduced by models that are fitted to the cluster data.
\end{enumerate}

The first concern above is straightforwardly addressed by jointly fitting or measuring all quantities of interest from the observations, and propagating the measurement errors using Monte Carlo sampling. The distribution of the resulting samples automatically contains all the information about the measurement covariance.

Unlike the first, the second and third issues results from decisions on the part of the observer. For the second issue noted above, the use of Monte Carlo sampling to handle the error propagation can again allow the measurement covariance to be straightforwardly understood.

In this paper, we are concerned with the third issue identified above, in particular as it applies to mass measurements of galaxy clusters based on the assumption of hydrostatic equilibrium (HSE), and scaling relations that are formed with such masses. We argue below that some widely employed procedures used to model cluster data introduce strong priors that can influence the resulting scaling relation constraints, and thus hamper our ability to perform robust astrophysical and cosmological measurements. Fortunately, there are simple, alternative approaches that do not suffer from these problems, which are also discussed here.

The paper is organized as follows. \secref{sec:background} provides a brief introduction to galaxy cluster mass measurements, scaling relations and the self-similar model. In \secref{sec:methods}, we discuss the various methods for estimating hydrostatic masses that have been proposed, with particular emphasis on the priors that each methods imposes on both masses and the resulting scaling relations. In \secref{sec:meta}, we review and discuss results on the mass--temperature relation slope from the literature, with attention to the impact of these modeling priors. Our conclusions are summarized in \secref{sec:summary}.

\section{Background} \label{sec:background}

\subsection{Hydrostatic mass estimates and scaling relations}

Many galaxy cluster mass estimates in the literature are based on X-ray observations.\footnote{While we focus on X-ray methodology and results, the central aspects of our discussion also apply to mass estimates based on other data such as the Sunyaev-Zel'dovich effect, galaxy number density or velocity dispersion.} X-ray data provide two observables that scale physically with total mass, namely the luminosity in the observed energy band and the temperature of the X-ray emitting, hot intracluster medium (ICM). Under the assumption of spherical symmetry, spectral and surface brightness data measured in projection can be de-projected, yielding three-dimensional profiles of emissivity and temperature. These two can be combined to infer the ICM density profile,\footnote{For the typical case of hot ($kT \gtsim 3\keV$), low-redshift clusters, and luminosity measured in the soft X-ray band (e.g. 0.5--2.0\keV{}), this conversion is essentially independent of temperature.} and thus the gas mass, as well as the bolometric luminosity \citep[e.g.][]{Sarazin1988xrec.book.....S}.

For clusters that are approximately spherical and close to HSE, such data can also be used to constrain the total mass profile. Specifically, HSE implies a relationship between the density and temperature profiles of the ICM, $n(r)$ and $T(r)$, and the total mass,
\begin{equation} \label{eq:hse}
 M(r) = -\frac{kT(r) r}{\mu\mproton G} \left( \frac{d \ln n}{d \ln r} + \frac{d \ln T}{d \ln r} \right),
\end{equation}
where $k$ is Boltzmann's constant, $G$ is Newton's constant, and $\mu\mproton$ is the mean molecular weight.

Conventionally, scaling relations are formed by relating the observables of interest to the total mass, $M_\Delta$, within a particular radius, $r_\Delta$, jointly defined by
\begin{equation} \label{eq:radius}
  M_\Delta = \frac{4\pi}{3} \Delta \rhocr(z) r_\Delta^3,
\end{equation}
where $\rhocr(z)$ is the critical density of the universe at the cluster's redshift. Typical choices for $\Delta$ range from 2500 (intermediate radius) to 200 (approximately the virial radius). Combining these equations, one can immediately write
\begin{eqnarray} \label{eq:rMsolution}
  r_\Delta^2 &=& \left[\frac{4\pi}{3}\Delta\rhocr(z)\right]^{-1} \left[\frac{k}{\mu\mproton G}\right] T(r_\Delta) \, \mathcal{F}(r_\Delta), \\
  M_\Delta &=& \left[\frac{4\pi}{3}\Delta\rhocr(z)\right]^{-1/2} \left[\frac{k}{\mu\mproton G}\right]^{3/2} T(r_\Delta)^{3/2} \, \mathcal{F}(r_\Delta)^{3/2}, \nonumber
\end{eqnarray}
where
\begin{equation}
  \mathcal{F}(r) = -\left( \frac{d \ln n}{d \ln r} + \frac{d \ln T}{d \ln r} \right).
\end{equation}

Clusters forming from idealized, spherical gravitational collapse with no additional heating or cooling are expected to have self-similar (i.e. described by the same function of $r/r_\Delta$ for every cluster) gas and dark matter density profiles. Assuming HSE, the temperature profiles of clusters will also have a self-similar shape (though not a common normalization). This case was studied by \citet{Kaiser1986MNRAS.222..323K}, who derived power-law predictions for scaling relations using masses defined by \eqnref{eq:radius}:\footnote{Often the factors $\rhocr(z)^{1/2}$ are written in terms of $E(z)$, the normalized Hubble parameter.}
\begin{eqnarray}
  \label{eq:selfsimilar}
  \Mgas{}_{,\Delta} & \propto & M_\Delta, \nonumber \\
  T_\Delta & \propto & \left[ \rhocr(z)^{1/2}M_\Delta \right]^{2/3}, \nonumber \\
  \rhocr(z)^{1/2}\,Y_\Delta & \propto & \left[ \rhocr(z)^{1/2}M_\Delta \right]^{5/3}, \nonumber \\
  \frac{L_\Delta}{\rhocr(z)^{1/2}} & \propto & \left[ \rhocr(z)^{1/2}M_\Delta \right]^{4/3}.
\end{eqnarray}
A constant gas mass fraction (the first line) is a direct consequence of the self-similar hypothesis, while the second follows from \eqnref{eq:rMsolution}, since self-similarity implies that $\mathcal{F}(r_\Delta)$ is a constant. Here $Y \propto \int dV \, n \, kT$ is the integrated, intrinsic Sunyaev-Zel'dovich signal (i.e. the thermal energy of the gas), and $L$ refers to the bremsstrahlung luminosity of the plasma ($L \propto \int dV \, n^2 T^{1/2}$). $T_\Delta$ may be the temperature at radius $r_\Delta$ or some weighted average of $T(r)$ within $r_\Delta$, since the scalings are identical given self similarity [$T(r_\Delta)/T_\Delta$ is constant].

For simplicity, we will henceforth eliminate most of the constants, setting
\begin{equation}
  \frac{k}{\mu \mproton G} = \frac{4\pi}{3} \Delta \rhocr(z) = 1.
\end{equation}
In practice, the redshift dependence represented by $\rhocr(z)$ must be properly accounted for; however, it is incidental to the focus of this work. By eliminating these terms, we effectively consider the simplified case of scaling relations at a single redshift and fixed density contrast.

\subsection{A simple example: the isothermal $\beta$ model} \label{sec:isothermalbeta}

As an example, we consider the isothermal $\beta$ model. This case is deliberately simplistic (indeed, the results below are well known), but it serves to illustrate some features of HSE mass estimation using parametrized models that are relevant to our discussion in \secref{sec:methods}.

In this model, the three-dimensional gas density and temperature profiles are parametrized by
\begin{eqnarray}
  n(r) &=& n_0\left(1+\frac{r^2}{\rc^2}\right)^{-3\beta/2}, \nonumber \\
  T(r) &=& T_0.
\end{eqnarray}
The gas mass is given by
\begin{eqnarray} \label{eq:betaMgas}
 \Mgas(r) & = & \frac{4\pi n_0 r^3}{3} ~ {}_2F_1\left( \frac{3}{2}, \frac{3}{2}\beta; \frac{5}{2}; -\frac{r^2}{\rc^2} \right),
\end{eqnarray} 
where ${}_2F_1(a,b;c;z)$ is the Gauss hypergeometric function.
 
Applying the hydrostatic equation, the mass profile is
\begin{eqnarray} \label{eq:beta}
  M(r) & = & 3 \beta T_0 r^3 \left(\frac{1}{r^2 + \rc^2}\right),
\end{eqnarray}
which yields a solution for the characteristic radius,
\begin{eqnarray} \label{eq:isothbetardelta}
  r_\Delta & = & \sqrt{3 \beta T_0 - \rc^2}.
\end{eqnarray}
From this, we can write the relationship between temperature and mass for clusters described by this model,
\begin{equation} \label{eq:betaTM}
  T_0 = \frac{1+\left(\rc/r_\Delta\right)^2}{3\beta} \, M_\Delta^{2/3}.
\end{equation}

In the self-similar case, all clusters have the same values of $\beta$ and $\rc/r_\Delta$, and so the self-similar scaling $M_\Delta \propto T_0^{3/2}$ follows directly from \eqnref{eq:betaTM}. Furthermore, the hypergeometric function in $\Mgas(r_\Delta)$ (\eqnref{eq:betaMgas}) assumes a constant value, leading to a constant gas mass fraction, and the other scaling laws in \eqnref{eq:selfsimilar} follow straightforwardly:
\begin{eqnarray}
  \Mgas{}_{,\Delta} \propto & r_\Delta^3 & \propto M_\Delta, \nonumber \\
  Y_\Delta \propto & \Mgas{}_{,\Delta} T_0 & \propto M_\Delta^{5/3}, \nonumber \\
  L_\Delta \propto & r_\Delta^3 T_0^{1/2} & \propto M_\Delta^{4/3}.
\end{eqnarray}

\begin{figure*}
  \centering
  \includegraphics[scale=\figscale]{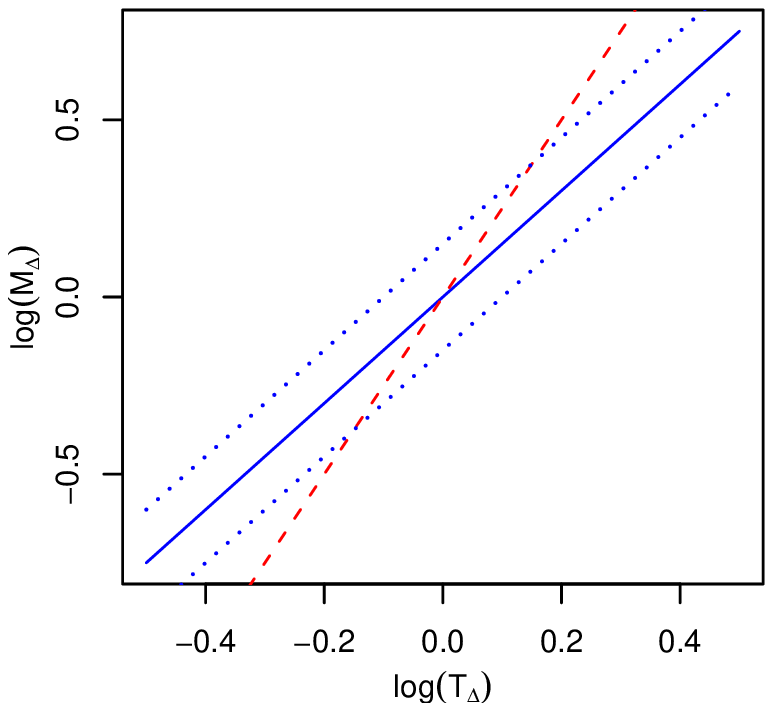}
  \hspace{5mm}
  \includegraphics[scale=\figscale]{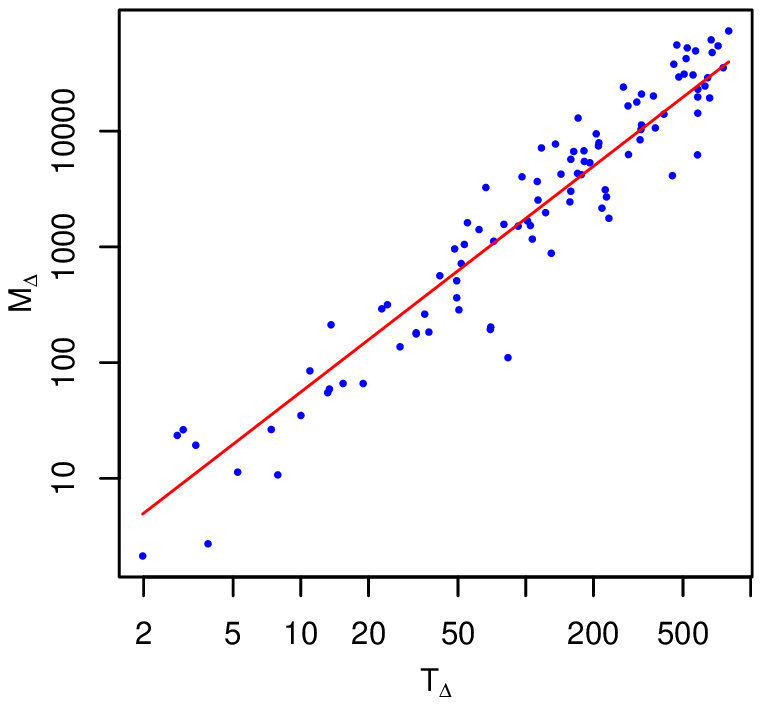}
  \caption{
    Left: Illustration of some of the simple scaling behaviors available to fully parametrized cluster models such as the isothermal $\beta$ model. The solid, blue line represents an exactly self-similar scaling relation, which is expected when the model is too restrictive (does not permit departures from self-similarity). If parameters that break self-similarity vary completely at random (e.g. if they are unconstrained by data), then the mean relation is still self-similar, with a scatter determined by the range over which the model parameters can vary (dotted, blue lines; see also right panel). However, if there is a systematic trend of the model parameters with mass or temperature, the slope of mean scaling relation can be perturbed (dashed, red line), and in general the relation need not be a power-law.
    Right: Numerical demonstration (see \appref{sec:sims}) of a self-similar scaling relation arising from random realizations of parametrized density and (non-isothermal) temperature profiles (corresponding to the case illustrated by dotted lines in the left panel). Here $T_\Delta$ is the emission-weighted projected temperature within $r_\Delta$. Although the model parameters are chosen randomly, and because their randomization is not mass dependent, the structure of the hydrostatic equation results in a self-similar mean scaling relation (red line). \appref{sec:sims} contains details of the parametrized models used and the allowed ranges for the model parameters.
  }
  \label{fig:simple}
\end{figure*}

Conversely, departures from self-similarity result in changes to these scaling laws. For example, consider the case in which $\rc/r_\Delta$ remains constant, but $\beta$ varies from cluster to cluster. The expectation value of the characteristic mass at fixed temperature can be written as
\begin{eqnarray} \label{eq:betaMTexpect}
  \expectation{M_\Delta|T_0} &=& \int_{-\infty}^\infty d\beta\, P(\beta|T_0) \left[\frac{3 \beta T_0}{1+\left(\rc/r_\Delta\right)^2}\right]^{3/2}, \nonumber\\
  & = & \expectation{\left.\beta^{3/2} \right| T_0} \left[\frac{3T_0}{1+\left(\rc/r_\Delta\right)^2}\right]^{3/2},
\end{eqnarray}
where $P(\beta|T_0)$ is the distribution of $\beta$ values for clusters with temperature $T_0$. Thus, if $\beta$ varies systematically with temperature as $\expectation{\left.\beta^{3/2}\right|T_0} \proptosim T_0^\alpha$, then the $M$--$T$ slope implied by this model is modified to $3/2+\alpha$ (\figref{fig:simple}). The generalization when both $\beta$ and $\rc/r_\Delta$ vary is straightforward, and the effect on the other scaling laws can be derived similarly.

\section{Approaches to Measuring Hydrostatic Masses} \label{sec:methods}

\subsection{Fully parametric} \label{sec:fullparametric}

When deriving hydrostatic cluster masses, a common practice is to fit parametric functions for the three-dimensional gas density and temperature profiles to the observed surface brightness and temperature data, and then to derive the total cluster mass profile using \eqnref{eq:hse}. For comparison with the other methods described below, it should be noted that selecting parametrized models for $n(r)$ and $T(r)$ is completely equivalent (via \eqnref{eq:hse}) to choosing parametrizations for $M(r)$ and $T(r)$, and thus implicitly imposes a prior on the form of $M(r)$. Because these functions share parameters, varying model parameters produces covariance in $M$ and $T$. That is, the choice of parametrized models also constitutes an implicit prior on the scaling relations, as described below.

Generalizing \eqnref{eq:betaMTexpect}, we can write the mean mass--temperature relation resulting from fits to parametrized $n(r)$ and $T(r)$ models as (\eqnref{eq:rMsolution})
\begin{eqnarray} \label{eq:paramMT}
  \expectation{M_\Delta|T_\Delta} &=& \int d\theta \, P(\theta|T_\Delta) \, T(r_\Delta; \theta)^{3/2} \, \mathcal{F}(r_\Delta; \theta)^{3/2}, \nonumber\\
  & = & T_\Delta^{3/2} \, \expectation{\left. \frac{T(r_\Delta)^{3/2}}{T_\Delta^{3/2}} \mathcal{F}(r_\Delta)^{3/2}\right|T_\Delta},
\end{eqnarray}
where $\theta$ represents the full set of parameters describing $n(r)$ and $T(r)$, and $T_\Delta$ is the measured gas temperature used to form the scaling relation. Typically, $T_\Delta$ is an emission-weighted average, dominated by the signal from relatively small radii, in which case $T(r_\Delta)/T_\Delta$ is a measure of the overall shape of the temperature profile. For self-similar clusters, both $T(r_\Delta)/T_\Delta$ and $\mathcal{F}(r_\Delta)$ are constant,\footnote{In principle, instrument-specific effects might make the ratio of $T(r_\Delta)$ to {\it measured} $T_\Delta$ vary, even for self-similar clusters. We do not consider such effects here.} and the self-similar slope of $3/2$ is trivially recovered. However, if either of these quantities varies systematically with mass, the slope may be perturbed from the self-similar value.

The explicit appearance of the exponent $3/2$ in \eqnref{eq:paramMT} makes clear that, in practice, our ability to detect departures from self-similarity using this approach depends on measuring the shape of the temperature and density profiles at $r_\Delta$. In the case of temperature, this is a challenging task for current X-ray observatories at even intermediate cluster radii (e.g. $\Delta=500$, a common choice). Furthermore, and not incidentally, the priors on the forms of the temperature and density (or mass, equivalently) profiles must be flexible enough to admit departures from self-similarity. In practice, the parametrizations employed are generally motivated by observable features of the surface brightness and temperature profiles, raising the possibility that the relatively low signal-to-noise at intermediate radii, and subsequent assumption of regular behavior in the profiles [i.e. similar values of $T(r_\Delta)/T_\Delta$ and $\mathcal{F}(r_\Delta)$], produces a bias favoring self-similarity.

Conversely, if the parametrizations provide too much flexibility near $r_\Delta$ to be effectively constrained by the data, then departures from self-similarity cannot be constrained either. In the extreme case where the data provide no constraint at all on the profiles near $r_\Delta$, the bracketed expression in \eqnref{eq:paramMT} simply samples the prior (the allowed region in model parameter space). If that prior is independent of $T_\Delta$, as common practice would dictate, the resulting scaling relation must have the self-similar slope on average, with the prior simply determining the size of the scatter about the relation. This behavior is explicitly demonstrated in right panel of \figref{fig:simple}, which shows a mass--temperature relation resulting from random realizations of a non-isothermal, parametrized cluster model (see \appref{sec:sims}). The randomized density and temperature profile models vary widely in shape and normalization, but because these variations have no mass dependence, the structure of \eqnref{eq:paramMT} results in a self-similar mean scaling relation.

Thus, the inability of current observatories to constrain high-resolution temperature profiles at the radii of interest poses a dilemma for the fully parametric approach to mass estimation. Allowing too little freedom in the adopted forms of the $n(r)$ and $T(r)$ profiles risks assuming implicitly that clusters are self-similar. On the other hand, allowing too much freedom can result in the profiles at $r_\Delta$ being so poorly constrained that departures from self-similarity in individual clusters cannot be constrained either; based on the argument above, this case may well also result in an apparently self-similar scaling relation on average.

Apart from temperature, the fully parametric mass estimate depends on the shape of $n(r)$ at $r_\Delta$. Both $\Mgas$ and $Y$ have a dependence on this quantity, being integrals of $n(r)$ weighted towards large radii. However, the surface brightness profile can be determined at much higher resolution than temperature from X-ray data, meaning that priors on the shape of $n(r)$ need not be as influential. Provided that the choice of density parametrization is not overly restrictive, we would thus expect biases towards self similarity to be less of a concern for the $\Mgas$--$M$ relation compared to $T$--$M$ or $Y$--$M$. The X-ray luminosity--mass relation should be essentially free of this bias, since it is dominated by emission from the dense gas at cluster centers, at radii typically $\ll r_\Delta$. It is therefore interesting to note that the $L$--$M$ relation is the only one of these scalings for which the fully parametric approach to mass estimation has consistently measured strong departures from self-similarity in the slope (e.g. \citealt{Vikhlinin0805.2207}; see also \secref{sec:meta}).

\subsection{Semi-parametric}

An alternative method for determining cluster hydrostatic mass profiles was developed by \citet[][see also \citealt*{White1997MNRAS.292..419W,Allen0110610,Schmidt0107311}]{Fabian1981ApJ...248...47F}. In this approach, a functional form for the total mass profile is explicitly adopted, and used in conjunction with a non-parametric description of the surface brightness to predict the temperature in concentric shells. Temperature measurements from spectral data then provide the means to constrain the parameters of the mass model.

In contrast to fully parametric methods, the semi-parametric approach does not restrict the forms of the ICM density or temperature profiles. Apart from the regularization imposed by the size of the annular regions analyzed (a factor in all of the approaches discussed here), these profiles are not constrained a priori. Whereas the fully parametric approach implies an implicit prior on the form of $M(r)$, the semi-parametric method explicitly adopts a prior on the form of this function, typically motivated by numerical simulations of cluster formation (e.g. the model of \citealt*{Navarro9611107}, hereafter \NFW{}).

As a consequence, the semi-parametric approach does not impose a prior on the form of the mass--temperature relation (or the other scaling relations) in the sense of \eqnref{eq:paramMT}. That is not to say that the procedure imposes no priors at all; the choice of a particular form for the mass profile explicitly does so. In this case, however, the effect of the prior on the mass reconstruction is completely transparent, and the goodness of fit furthermore provides a means to evaluate the mass model, a significant advantage.

\subsection{Non-parametric}

The most general possibility for hydrostatic mass analysis is a fully non-parametric de-projection. Examples include the methods introduced by \citet{Arabadjis0305547} and \citet{Ameglio0811.2199}, in which numerical derivatives of non-parametric ICM density and temperature profiles are directly used to reconstruct the enclosed mass, subject to the constraint that mass increase with radius; and by \citet{Nulsen1008.2393}, in which the total densities in concentric spherical shells are free model parameters. In a sense, these approaches are, respectively, logical extensions of the fully parametric methods, which use the derivatives of $n(r)$ and $T(r)$ to derive the mass, and the semi-parametric methods, which model the mass profile directly. However, these non-parametric methods require very high quality data compared to methods which impose some kind of prior on the mass distribution; as has already been mentioned, X-ray data typically cannot resolve the temperature gradient near $r_{500}$. The use of these approaches has thus been relatively limited.

\subsection{Non-hydrostatic proxies}

Finally, the explicit assumption of HSE can be bypassed by estimating mass using a proxy (e.g. $\Mgas$ or $\Yx=\Mgas T_\Delta$) from an external scaling relation. This approach clearly carries its own prior, namely the validity of the mass proxy, which must be verified and calibrated using true mass determinations. There are also restrictions on what scaling relations can sensibly be investigated using this technique; for example, given its definition, $\Yx$-derived masses should not be used to investigate scalings with gas mass or temperature. On the other hand, for hot clusters ($kT\gtsim4\keV$), $\Mgas$ is a good mass proxy whose determination is essentially independent of temperature \citep{Allen0706.0033}, so masses estimated from $\Mgas$ can reasonably be used to study the $M$--$T$ relation in this mass range (e.g. \citealt{Mantz0909.3099}). The appropriate use of mass proxies can thus potentially increase the available sample size and redshift range for studying some scaling relations.

\section{Meta-analysis of Constraints on the Mass--Temperature Slope} \label{sec:meta}

The comparison of scaling relations derived in different works is complicated by a variety of potential systematics, including (potentially redshift-dependent) selection effects, instrument cross-calibration, and the use of different regression methods over the years, in addition to the issues discussed in this paper (see also \appref{sec:comments}). Nevertheless, it is interesting to test whether there is any trend in scaling relation results from the literature with the mass modeling technique employed.

\begin{figure}
  \centering
  \includegraphics[scale=\figscale]{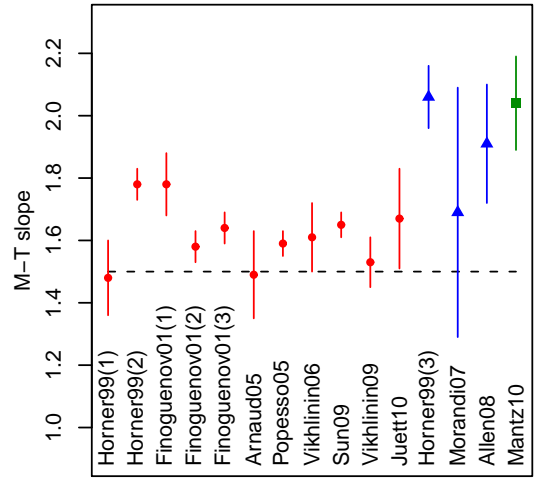}
  \caption{Mass--temperature slopes and 68.3 per cent uncertainties from the literature (see text for citations). In some cases, we have included multiple results from the same paper, based on different data sets or mass models. Red circles indicate results obtained from fitting parametric $n(r)$ and $T(r)$ models; blue triangles show results using a non-parametric description of the ICM along with an explicit prior on the form of the total mass profile; and the green square uses gas mass as a proxy for total mass. The dashed line shows the self-similar value of the slope.}
  \label{fig:MTslopes}
\end{figure}

Here we focus on mass--temperature relations measured from X-ray data, which have a particularly long history. A sampling of $M$--$T$ slopes and reported uncertainties from the literature over the past 12 years is shown in \figref{fig:MTslopes} \citep{Horner9902151,Finoguenov0010190,Arnaud0502210,Popesso0411536,Vikhlinin0507092,Vikhlinin0805.2207,Morandi0704.2678,Allen0706.0033,Sun0805.2320,Juett0912.4078,Mantz0909.3099}. In some cases, we have included multiple results from the same authors, where different data sets or mass models produced noticeably different results. In the figure, red circles are results obtained by fitting fully parametric $n(r)$ and $T(r)$ models. Blue triangles indicate results using a non-parametric description of the ICM along with an explicit prior on the form of the total mass profile (semi-parametric methods). The green square reflects a study of massive clusters where gas mass was used as a proxy for total mass. Temperature measurements used in the displayed results are all emission-weighted averages, and masses in a given study are estimated at a constant value of $\Delta$ (see below).

Among the results based on fits to very simple $n(r)$ and $T(r)$ models are:
\begin{enumerate}
\item Two results from \citet{Horner9902151} where masses were estimated from X-ray observations. The first slope, $1.48\pm0.12$, is from a heterogeneous sample with measured temperature profiles. For the second, the isothermal $\beta$ model was fitted to the data of \citet{Fukazawa1997PhDT........24F}, resulting in a slope of $1.78\pm0.05$. In both cases (as well as their third result, below), masses were rescaled to $\Delta=200$.

\item The first two results from \citet{Finoguenov0010190} use a compilation of clusters for which resolved temperature profiles were available, with masses estimated at $\Delta=500$ by fitting a $\beta$ model density profile and assuming a polytropic relationship between gas density and temperature. The first slope, $1.78\pm0.10$, uses the entire sample, while the second, $1.58\pm0.05$, was obtained by excluding 4 clusters with measured $\beta<0.4$.

\item The third result shown from \citet{Finoguenov0010190} and the slope from \citet{Popesso0411536} were derived by fitting different subsets of the data of \citet{Reiprich0111285}, respectively finding slopes of $1.64\pm0.05$ and $1.59\pm0.04$. The \citet{Reiprich0111285} analysis provides masses at $\Delta=500$ from isothermal $\beta$ model fits.
\end{enumerate}

More recent works using $n(r)$ and $T(r)$ fits generally have used more complicated models, which are detailed in the respective papers:
\begin{enumerate}
\item \citet{Arnaud0502210} used masses measured by \citet{Pointecouteau0501635}, who fitted functions for $n(r)$ and $T(r)$ to X-ray data. The figure shows their slope of $1.49\pm0.14$ for clusters with $kT>3.5\keV$ at $\Delta=1000$, the largest radius for which no extrapolation was required. Their results at other radii are very similar.

\item \citet{Vikhlinin0507092} fitted parametrized models to 13 clusters, obtaining mass--temperature relations at $\Delta=2500$ and 500 with both emission- and mass-weighted temperatures. In the figure, we show the emission-weighted slope for $\Delta=500$, $1.61\pm0.11$. The best fitting values in the other cases ranged from 1.51 to 1.64. The analysis was extended to 17 clusters in \citet{Vikhlinin0805.2207}, resulting in a slope of $1.53\pm0.08$.

\item \citet{Sun0805.2320} fitted a sample of 23 groups and 14 clusters, spanning $0.7\keV<kT<11\keV$, obtaining a slope of $1.65\pm0.04$ at $\Delta=500$.

\item \citet{Juett0912.4078} fitted the models of \citet{Vikhlinin0507092} to 28 clusters with $kT>2\keV$, finding a slope of $1.67\pm0.16$ at $\Delta=500$.
\end{enumerate}

The mass--temperature slopes that rely on fully parametric fits to $n(r)$ and $T(r)$ tend to cluster in the 1.50--1.65 range. The exceptions are one result from \citet[][using the isothermal $\beta$ model]{Horner9902151} and one from \citet[][using a polytropic model]{Finoguenov0010190}. In the former case, \citet{Horner9902151} comment that there exists a clear correlation between measured values of $T_0$ and $\beta$ in the data \citep{Fukazawa1997PhDT........24F}, $\beta \propto T_0^{0.26 \pm 0.03}$. Based on \secref{sec:isothermalbeta}, one might expect such a correlation to result in a steeper slope when the isothermal $\beta$ model is used. The simple $3/2+\alpha$ formula from \secref{sec:isothermalbeta} over-predicts the size of the effect: $\beta^{3/2} \proptosim T^{0.39}$ implies a yet steeper slope than was observed. The full explanation likely involves the effect of the third fit parameter, $\rc$, as well as the measurement errors and the method used to fit the scaling relation. In \citet{Finoguenov0010190}, eliminating the clusters with the smallest $\beta$ measurements (which also happen to be at the low-temperature end of the data set) reduces both the empirical $\beta$--$T_0$ correlation and the mass--temperature slope (compare their first and second results in the figure), supporting the qualitative notion that model parameter correlations contribute to steepening of the slope. Similarly, the \citet{Reiprich0111285} data set used by \citet[][their third result above]{Finoguenov0010190} and \citet{Popesso0411536} has an empirically smaller correlation between $T_0$ and $\beta$, and fits to a correspondingly shallower slope. The works using more complicated $n(r)$ and $T(r)$ models generally show less strong departures from the self-similar value.

Relatively fewer authors have used an explicit prior on the form of the mass profile, along with a non-parametric description of the ICM:
\begin{enumerate}
\item The third result from \citet{Horner9902151} employs masses from \citet{White1997MNRAS.292..419W}, obtaining a slope of $2.06\pm0.10$. The mass profiles were constrained by a combination of galaxy velocity dispersion and X-ray temperature data.

\item \citet{Morandi0704.2678} fitted a mass profile motivated by \citet{Rasia0309405} to X-ray data for 24 hot ($kT>5\keV$) clusters, obtaining a slope of $1.7\pm0.4$ at $\Delta=2500$. However, when they allow the normalization of the scaling relation to evolve with redshift, the measured mass--temperature slope is steeper, $2.30\pm0.24$.

\item \citet{Allen0706.0033} fitted an \NFW{} mass profile to non-parametric surface brightness and temperature data for 42 massive, dynamically relaxed clusters, obtaining HSE masses at $\Delta=2500$. The \NFW{} profile provides an acceptable fit to the data (see also \citealt{Schmidt0610038}). Combining these mass measurements with temperatures from an extension of the work in \citet{Mantz0909.3099}, we obtain a mass--temperature slope of $1.91\pm0.19$ (see \appref{sec:A08MT}).
\end{enumerate}

The final result shown in the figure is from \citet{Mantz0909.3099}, who used gas mass as a proxy to estimate total masses at $\Delta =500$ for a sample of 94 hot, massive clusters, obtaining a mass--temperature slope of $2.04\pm0.15$. Because the gas mass fraction was calibrated using the data of \citet{Allen0706.0033}, the two results are not entirely independent. On the other hand, relatively few of the \citet{Mantz0909.3099} clusters are in the \citet{Allen0706.0033} data set, so this dependence should largely be limited to the normalization of the scaling relation.

Apart from the \citet{Morandi0704.2678} slope, which has a large uncertainty, the results that use explicit priors on the form of the mass profile or employ gas mass as a proxy appear to prefer a relatively steep slope compared with the other works, $\sim 2.0$. Given that mass models such as the \NFW{} profile are well motivated by numerical simulations and provide an acceptable fit to cluster data (e.g. \citealt{Schmidt0610038}), the segregation apparent in \figref{fig:MTslopes} suggests that the implicit priors in fully parametric $n(r)$ and $T(r)$ models bias the resulting $M$--$T$ slopes towards the self-similar value.

\section{Summary} \label{sec:summary}

In this paper, we have discussed the influence of priors on hydrostatic mass estimates of clusters and on the resulting mass--observable scaling relations. The use of fully parametric gas density (or X-ray brightness) and temperature profiles, similar to those commonly used in the literature, introduces an implicit prior on the form of the mass profile via the hydrostatic equation. Furthermore, the structure of the prior thus imposed results in an implicit prior on the cluster scaling relations. If the parametrized models employed are insufficiently flexible, or conversely if they are too general to be constrained at the radii of interest, then constraints on the scaling relations will be biased towards having self-similar slopes.

Alternative techniques for hydrostatic mass measurement exist which, by construction, do not suffer from this bias. The most common of these is a semi-parametric approach, in which a parametric prior on the form for the mass profile is explicitly adopted, with the ICM described independently and non-parametrically. Typically, the priors used here are motivated by the results of numerical simulations. An advantage of the semi-parametric approach is that it requires no a priori assumptions about the potentially complex form of the ICM density and temperature profiles, and that the applicability of the mass profile model can be straightforwardly evaluated through the goodness of fit. We comment further on the relative merits of various methods, and offer general recommendations, in \appref{sec:comments}.

In the literature, results for the mass--temperature slope obtained by fitting parametric $n(r)$ and $T(r)$ profiles tend to cluster relatively near the self-similar value. Semi-parametric analyses appear to prefer a significantly steeper mass--temperature relation, although there are relatively few such works to consider. While a variety of systematic effects can potentially affect the scaling relations, this segregation of values for the $M$--$T$ slope suggests that the priors imposed during mass estimation have a significant influence that needs to be considered carefully.

As cluster surveys at all wavelengths are expanded to higher and higher redshifts, and are used to investigate more complex cosmological questions, accurate calibration of the relevant scaling relations will only become more important. Gravitational lensing will make a unique contribution to these efforts, particularly in assessing the residual bias in ICM-based mass estimates due to the HSE assumption. Nevertheless, ICM-based mass measurements for relaxed systems will remain an important ingredient in cluster cosmology due to the higher precision and lower systematic scatter of individual estimates compared to lensing. It is therefore critical, going forward, that the priors employed in these measurements be minimal, straightforwardly testable, and well understood.

\section*{Acknowledgments}

The authors are grateful to Mark Voit for interesting and insightful comments. AM was supported by an appointment to the NASA Postdoctoral Program at the Goddard Space Flight Center, administered by Oak Ridge Associated Universities through a contract with NASA. SWA acknowledges supported from the U.S. Department of Energy under contract number DE-AC02-76SF00515.

\def \aap {A\&A} 
\def \aapr {A\&AR} 
\def \statisci {Statis. Sci.} 
\def \physrep {Phys. Rep.} 
\def \pre {Phys.\ Rev.\ E} 
\def \sjos {Scand. J. Statis.} 
\def \jrssb {J. Roy. Statist. Soc. B} 
\def \pan {Phys. Atom. Nucl.} 
\def \epja {Eur. Phys. J. A} 
\def \epjc {Eur. Phys. J. C} 
\def \jcap {J. Cosmology Astropart. Phys.} 
\def \ijmpd {Int.\ J.\ Mod.\ Phys.\ D} 
\def \nar {New Astron. Rev.}

\def \araa {ARA\&A}
\def \aj {AJ}
\def \aar {A\&AR}
\def \apj {ApJ}
\def \apjl {ApJL}
\def \apjs {ApJS}
\def \mnras {MNRAS}
\def \nat {Nat}
\def \pasj {PASJ}
\def \science {Sci}

\def \gca {Geochim.\ Cosmochim.\ Acta}
\def \npa {Nucl.\ Phys.\ A}
\def \plb {Phys.\ Lett.\ B}
\def \prc {Phys.\ Rev.\ C}
\def \prd {Phys.\ Rev.\ D}
\def \prl {Phys.\ Rev.\ Lett.}

\appendix

\section{Simulations} \label{sec:sims}

As discussed in \secref{sec:fullparametric}, when the model parameters that determine the density and temperature profiles at $r_\Delta$ are unconstrained or poorly constrained, the fully parametric approach can be biased towards self-similar scaling relations. As an explicit illustration of this, consider a $\beta$ model description of the gas density in conjunction with a simple, non-isothermal temperature profile,
\begin{eqnarray}
 T(r) &=& \frac{T_0}{\left[1+(r/\rt)^b\right]^{c/b}}.
\end{eqnarray}
This function is a simplification of the form used by \citet{Vikhlinin0507092}, namely eliminating the `cool core' term, which is intended to describe the profile at small radii.

To illustrate the case where these models are effectively unconstrained, we generated random realizations by sampling independent, uniform values of the model parameters within the ranges given in \tabref{tab:nonisobetarand}. The radial scales in the density and temperature models, $\rc$ and $\rt$, were allowed to take values between zero and $\rmax = \mathrm{max} \sqrt{3\beta T_0}$. This is the maximum value of $\rc$ for which the isothermal $\beta$ model has a real solution for $r_\Delta$ (\eqnref{eq:isothbetardelta}); while the same is not true of this non-isothermal model, allowing larger values does not change the resulting picture qualitatively. $\beta$ was allowed to vary over a range somewhat wider than that seen in observations, while the temperature exponents, $b$ and $c$, were varied over approximately the range allowed by \citet{Vikhlinin0507092}. To provide an adequate baseline to observe the resulting scaling behavior, the temperature normalization, $T_0$, was sampled uniformly in the logarithm between 1 and 1000.

\begin{table}
 \centering
 \caption{Allowed ranges for uniformly distributed parameters of the non-isothermal $\beta$ model, with $\rmax$ defined as $\mathrm{max} \sqrt{3\beta T_0}$. The gas density normalization is not shown, since it has no effect on the mass--temperature relation.}
 \label{tab:nonisobetarand}
 \begin{tabular}{ccc}
   \hline
   Parameter & min & max   \\
   \hline
   $\beta$ & 0.1 & 1.3     \\
   $\rc$   &  0  & $\rmax$ \\
   $\log_{10}(T_0)$   &  0  & 3    \\
   $\rt$   &  0  & $\rmax$ \\
   $b$     &  0  & 5       \\
   $c$     &  0  & 10      \\
   \hline
 \end{tabular}
\end{table}

\begin{figure}
  \centering
  \includegraphics[scale=\figscale]{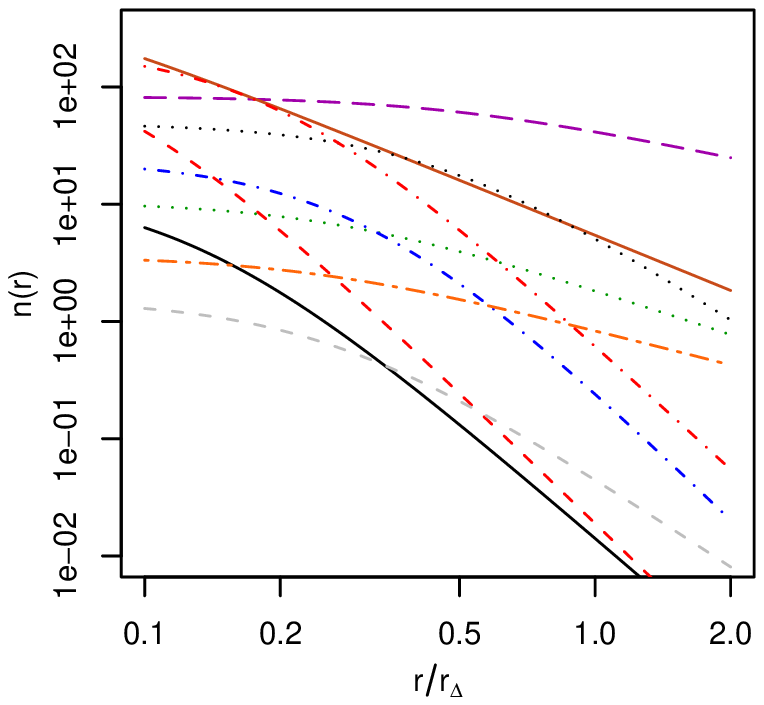}
  \includegraphics[scale=\figscale]{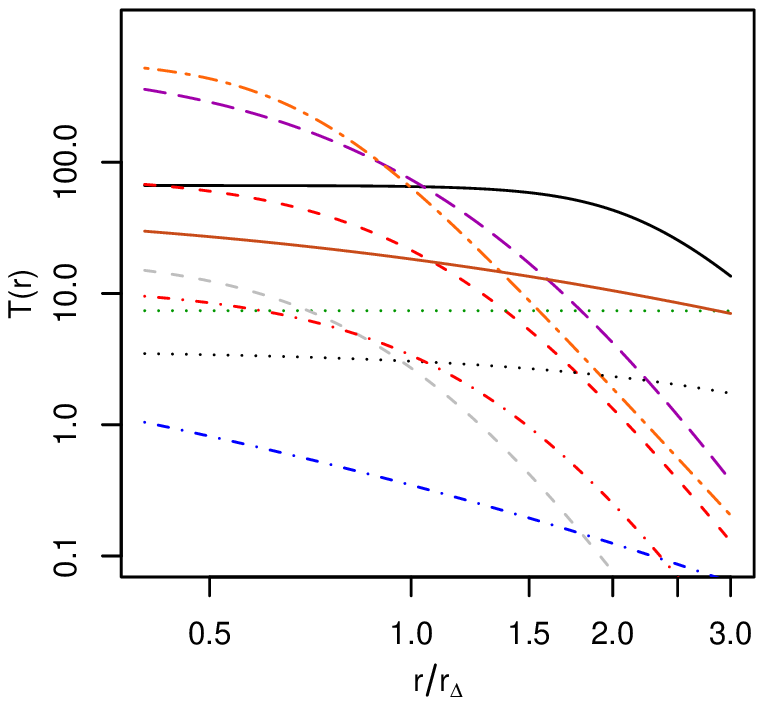}
 \caption{A sample of 10 randomized gas density and temperature profile models, demonstrating the range of behavior spanned by the randomization (\tabref{tab:nonisobetarand}).}
  \label{fig:nonisobetaprofiles}
\end{figure}

For each realization, an implicit solution for $r_\Delta$ (\eqnref{eq:rMsolution}) was searched for numerically, and models for which there was no real solution were discarded. A sample of the resulting density and temperature profiles is shown in \figref{fig:nonisobetaprofiles}. The model profiles are clearly not self-similar in any meaningful sense, but, because their variation is independent of mass, the slopes of the mean scaling relations take on the self-similar values (right panel of \figref{fig:simple}). For clarity, we have culled models where $\rc/r_\Delta>0.7$ from the figure; these models produce an asymmetric tail to low masses, but do not change the scaling relation slope. The x-axis of the figure shows the emission-weighted projected temperature within $r_\Delta$, calculated from the $n(r)$ and $T(r)$ profiles, although the precise definition of $T_\Delta$ does not affect the conclusions.

\section{Recommendations} \label{sec:comments}

Here we offer some brief thoughts on the task of obtaining hydrostatic mass estimates using minimal assumptions apart from hydrostatic equilibrium and spherical symmetry. Foremost, it must be emphasized that invoking HSE for systems that are not dynamically relaxed\footnote{In X-rays, dynamically relaxed systems are generally identified as having sharp surface brightness peaks and minimal isophote rotation or centroid variation.} results in significant bias and systematic scatter at the level of tens of per cent \citep*[e.g.][]{Nagai0609247}. For relaxed clusters and measurements made at intermediate radii ($r \sim r_{2500}$), bias and scatter due to residual departures from equilibrium should be at the $\sim 10$ per cent level or less. However, even in relaxed clusters, the assumption of HSE should be avoided in the outer regions ($r \gtsim  r_{500}$) where gas clumping \citep{Simionescu1102.2429} and increased non-thermal pressure support \citep{Nagai0609247,Pfrommer0611037,Mahdavi0710.4132} may occur. The same is true of the central few tens of kpc in systems where the influence of the central active galactic nucleus is often evident \citep[e.g.][]{Fabian0306036,Forman0312576,McNamara0709.2152}.

Ideally, a non-parametric method would be used to estimate masses, for example the model recently described by \citet{Nulsen1008.2393}. In this particular approach, the cluster is modeled as a series of concentric, spherical shells, with constant temperature and total (dark matter + baryons) density in each shell. These non-parametric temperature and mass profiles, with the addition of an overall gas density normalization and under the assumption of HSE, determine the gas density at all radii. Although this method is advantageous in principle, in practice it is only feasible with very high-quality data \citep{Nulsen1008.2393}.

More generally, some kind of regularization, in the form of an analytic model, is required to constrain hydrostatic masses. Based on the considerations discussed in this paper, it is preferable to apply this model to the total mass profile (e.g. the \NFW{} model) rather than the ICM. Considered as a modification of the above algorithm, the resulting semi-parametric model is parametrized by a set of temperatures in concentric shells, a normalization for the gas density, and the parameters of the chosen total mass model. Precisely this approach was used recently by \citet{Simionescu1102.2429}, who adopted an \NFW{} description of the mass profile to model \Suzaku{} data for the Perseus cluster.\footnote{The \citet{Nulsen1008.2393} model, including both non-parametric and NFW variants, is expected to be included in the next public release of {\tt XSPEC} (P. Nulsen, private communication).} The similar method of \citet{Fabian1981ApJ...248...47F}, in which the ICM is parametrized by the surface brightness in concentric annuli and a pressure normalization at large radius, has also found use in the literature \citep{White1997MNRAS.292..419W,Allen0110610,Allen0405340,Allen0706.0033,Schmidt0107311,Schmidt0610038,Siemiginowska1008.1739}. We note that the use of a parametrized mass profile along with the assumption of HSE typically allows the non-parametric temperature profile to be modeled at higher spatial resolution than in either a fully non-parametric mass solution or a simple, geometric de-projection (e.g. using the {\tt PROJCT} model in {\tt XSPEC}\footnote{\url{http://heasarc.gsfc.nasa.gov/docs/xanadu/xspec/}}) where the mass is not modeled at all.

For fitting the scaling relations themselves, we note that a full treatment generically requires simultaneous modeling of the cluster mass function due to selection effects \citep{Mantz0909.3098,Allen1103.4829}. Only when the intrinsic covariance between the observable of interest and the observable used to select the cluster sample is sufficiently small can approximate results be obtained without explicitly modeling the mass function and selection process. In this case, the analysis should still include a full treatment of heteroscedastic and possibly correlated measurement errors, and intrinsic scatter \citep[e.g.][]{Gelman2004BayesianDataAnalysis,Kelly0705.2774}.

\section{Allen et al. 2008 Mass--Temperature Relation} \label{sec:A08MT}

\begin{figure}
  \centering
  \includegraphics[scale=\figscale]{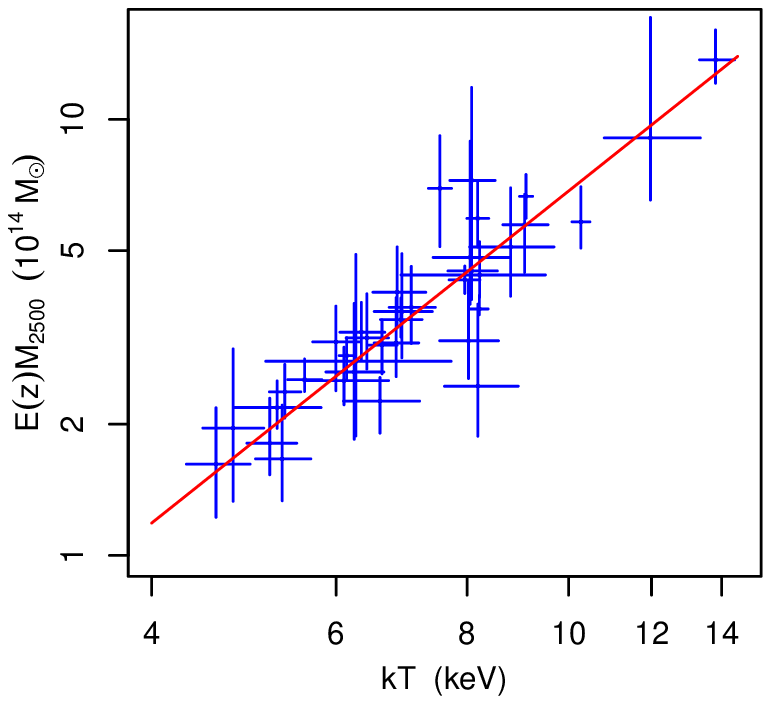}
  \caption{Scaling relation from the cluster mass measurements of \citet{Allen0706.0033} and temperature measurements of \citet{Mantz0909.3099}. The best fitting power law, shown by the red line, has a slope of $1.91\pm0.19$. The fit accounts for measurement errors in both quantities and is not sensitive to possible correlation of the measurement uncertainties for each cluster (see text).}
  \label{fig:A08MT}
\end{figure}

\citet{Allen0706.0033} used \Chandra{} X-ray Observatory data to measure hydrostatic masses at $r_{2500}$ for 42 hot ($kT>4.5\keV$), dynamically relaxed clusters at redshifts $0.05<z<1.1$. As described in \secref{sec:meta}, mass estimates were obtained by fitting an \NFW{} profile to the total mass distribution, using a non-parametric description of the ICM surface brightness and temperature. Average temperatures for some of these clusters, with a more current version of the \Chandra{} calibration, were measured by \citet{Mantz0909.3099}. We report here temperatures for many of the remaining clusters, from data reduced using exactly the same procedure as described in that paper. The temperatures are typically measured within larger radii than $r_{2500}$, but (being emission-weighted) the difference between these measurements and true $kT_{2500}$ values is at the few per cent level, typically smaller than the statistical error bars \citep[e.g.][]{Vikhlinin0805.2207,Mantz0909.3099}.\footnote{Specifically, temperatures were fit from spectra in annuli between $100\kpc$ and the radius for each cluster where the signal-to-noise of the 0.8--7.0\keV{} surface brightness profile falls to 2, as described in \citet{Mantz0909.3099}. These temperatures are a closer match spatially to $r_{2500}$, and are less prone to background modeling systematics, than the $kT_{500}$ values reported in that work.} These average temperatures are listed in \tabref{tab:A08MT}, along with the best fitting values and 68.3 per cent confidence intervals for $M_{2500}$ from \citet{Allen0706.0033}. We fit a power-law to the data using the {\tt linmix\_err} algorithm of \citet{Kelly0705.2774}, assuming independent, log-normal measurement errors.\footnote{The assumption of independent error bars violates our own advice from \secref{sec:intro}. We have explicitly verified that measurement error correlations of $\pm 0.9$ produce negligible change to the best fitting power-law slope in this case.} The resulting best fit, shown in \figref{fig:A08MT}, has a slope of $1.91\pm0.19$.

\renewcommand{\arraystretch}{1.3}
\begin{table*}
  \centering
  \caption{Redshifts, masses at $r_{2500}$, and average temperatures for \citet{Allen0706.0033} clusters. Mass constraints correspond to the \NFW{} fits reported in \citet{Allen0706.0033}, assuming a spatially flat, cosmological constant model with Hubble parameter $H_0=70\km\second^{-1}\Mpc^{-1}$ and mean matter density with respect to the critical value $\Omegam=0.3$. Values of the normalized Hubble parameter at each cluster's redshift, $E(z) = H(z)/H_0 \propto \rhocr(z)^{1/2}$, are given for this cosmology. Temperatures were derived as described in \citet{Mantz0909.3099}, apart from Abell clusters 1795, 2029 and 478, whose temperatures are from \citet{Horner9902151}. To prevent the temperature data from becoming too heterogeneous, the five clusters which were not studied in either of these works were omitted from the analysis. (We have kept the Horner et al. temperatures for consistency with \citealt{Mantz0909.3099}, where they were also used; however, they do not influence the fit significantly.)}
  \label{tab:A08MT}
  {\footnotesize
  \begin{tabular}{lcccclcccc}
    \hline
    Name & $z$ & $E(z)$ & $M_{2500}$ & $kT$ & Name & $z$ & $E(z)$ & $M_{2500}$ & $kT$\vspace{-1.5mm} \\
    & & & $(10^{14}\Msun)$ & $(\keV)$ & & & & $(10^{14}\Msun)$ & $(\keV)$ \\ 
    \hline

Abell~1795          &  0.063  &  1.030  &  $2.79_{-0.34}^{+0.28}$  &  $6.14_{-0.10}^{+0.10}$  &  MACSJ1931.8$-$2635  &  0.352  &  1.201  &  $4.02_{-0.88}^{+3.41}$  &  $8.05_{-0.63}^{+0.75}$  \\
Abell~2029          &  0.078  &  1.037  &  $3.54_{-0.11}^{+0.26}$  &  $8.22_{-0.16}^{+0.16}$  &  MACSJ1115.8$+$0129  &  0.355  &  1.203  &  $6.02_{-2.49}^{+3.82}$  &  $8.08_{-0.38}^{+0.43}$  \\
Abell~478           &  0.088  &  1.042  &  $4.11_{-0.29}^{+0.31}$  &  $7.96_{-0.27}^{+0.27}$  &  MACSJ1532.9$+$3021  &  0.363  &  1.208  &  $3.32_{-0.57}^{+0.90}$  &  $6.86_{-0.36}^{+0.44}$  \\
PKS0745$-$191       &  0.103  &  1.050  &  $4.97_{-0.84}^{+0.98}$  &  ---                     &  MACSJ0011.7$-$1523  &  0.378  &  1.219  &  $2.59_{-0.39}^{+0.68}$  &  $6.42_{-0.29}^{+0.31}$  \\
Abell~1413          &  0.143  &  1.071  &  $3.51_{-0.33}^{+0.31}$  &  ---                     &  MACSJ1720.3$+$3536  &  0.391  &  1.228  &  $3.02_{-0.53}^{+0.73}$  &  $7.08_{-0.33}^{+0.38}$  \\
Abell~2204          &  0.152  &  1.076  &  $4.09_{-0.45}^{+0.79}$  &  $8.22_{-1.30}^{+1.28}$  &  MACSJ0429.6$-$0253  &  0.399  &  1.234  &  $1.83_{-0.29}^{+0.24}$  &  $6.61_{-0.51}^{+0.60}$  \\
Abell~383           &  0.188  &  1.097  &  $2.16_{-0.28}^{+0.34}$  &  $5.36_{-0.18}^{+0.19}$  &  MACSJ0159.8$-$0849  &  0.404  &  1.237  &  $4.63_{-1.03}^{+0.81}$  &  $9.08_{-0.42}^{+0.48}$  \\
Abell~963           &  0.206  &  1.107  &  $2.74_{-0.39}^{+0.39}$  &  $6.64_{-0.21}^{+0.22}$  &  MACSJ2046.0$-$3430  &  0.423  &  1.251  &  $1.57_{-0.50}^{+0.81}$  &  $4.78_{-0.31}^{+0.33}$  \\
RXJ0439.0$+$0521    &  0.208  &  1.109  &  $1.63_{-0.25}^{+0.44}$  &  $5.19_{-0.25}^{+0.32}$  &  MACSJ1359.2$-$1929  &  0.447  &  1.268  &  $2.20_{-0.72}^{+1.67}$  &  $6.27_{-1.13}^{+1.46}$  \\
RXJ1504.1$-$0248    &  0.215  &  1.113  &  $5.33_{-0.75}^{+1.12}$  &  $8.19_{-0.19}^{+0.20}$  &  MACSJ0329.7$-$0212  &  0.450  &  1.271  &  $2.56_{-0.35}^{+0.44}$  &  $6.34_{-0.29}^{+0.34}$  \\
Abell~2390          &  0.230  &  1.122  &  $5.19_{-0.67}^{+1.06}$  &  $10.2_{-0.20}^{+0.21}$  &  RXJ1347.5$-$1144    &  0.451  &  1.271  &  $10.8_{-1.26}^{+1.85}$  &  $13.8_{-0.48}^{+0.59}$  \\
RXJ2129.6$+$0005    &  0.235  &  1.125  &  $2.34_{-0.70}^{+1.02}$  &  $6.24_{-0.38}^{+0.42}$  &  3C295               &  0.461  &  1.279  &  $1.71_{-0.18}^{+0.26}$  &  $5.27_{-0.48}^{+0.53}$  \\
Abell~1835          &  0.252  &  1.135  &  $5.87_{-0.65}^{+0.72}$  &  $9.11_{-0.13}^{+0.13}$  &  MACSJ1621.6$+$3810  &  0.461  &  1.279  &  $2.83_{-0.62}^{+1.01}$  &  $6.93_{-0.41}^{+0.48}$  \\
Abell~611           &  0.288  &  1.158  &  $2.65_{-0.43}^{+0.72}$  &  $6.85_{-0.32}^{+0.35}$  &  MACSJ1427.3$+$4408  &  0.487  &  1.299  &  $1.88_{-0.44}^{+1.02}$  &  $8.19_{-0.58}^{+0.76}$  \\
Zwicky~3146         &  0.291  &  1.156  &  $5.99_{-1.59}^{+1.92}$  &  $7.54_{-0.19}^{+0.20}$  &  MACSJ1311.0$-$0311  &  0.494  &  1.304  &  $2.37_{-0.54}^{+0.49}$  &  $6.00_{-0.30}^{+0.33}$  \\
Abell~2537          &  0.295  &  1.163  &  $2.67_{-0.48}^{+0.98}$  &  $8.03_{-0.49}^{+0.55}$  &  MACSJ1423.8$+$2404  &  0.539  &  1.339  &  $2.59_{-0.23}^{+0.31}$  &  $6.92_{-0.30}^{+0.33}$  \\
MS2137.3$-$2353     &  0.313  &  1.174  &  $2.15_{-0.13}^{+0.25}$  &  $5.56_{-0.20}^{+0.22}$  &  MACSJ0744.9$+$3927  &  0.686  &  1.462  &  $3.07_{-0.43}^{+0.86}$  &  $8.08_{-0.41}^{+0.47}$  \\
MACSJ0242.6$-$2132  &  0.314  &  1.175  &  $2.14_{-0.26}^{+0.41}$  &  $6.10_{-0.52}^{+0.63}$  &  MS1137.5$+$6625     &  0.782  &  1.548  &  $2.81_{-0.77}^{+1.96}$  &  ---                     \\
MACSJ1427.6$-$2521  &  0.318  &  1.178  &  $1.37_{-0.34}^{+0.48}$  &  $4.61_{-0.29}^{+0.36}$  &  CLJ1226.9$+$3332    &  0.892  &  1.653  &  $5.46_{-1.54}^{+4.89}$  &  $11.9_{-1.16}^{+1.39}$  \\
MACSJ2229.8$-$2756  &  0.324  &  1.182  &  $1.41_{-0.28}^{+0.46}$  &  $5.33_{-0.30}^{+0.35}$  &  CLJ1415.2$+$3612    &  1.028  &  1.789  &  $0.97_{-0.23}^{+0.39}$  &  ---                     \\
MACSJ0947.2$+$7623  &  0.345  &  1.196  &  $4.26_{-0.98}^{+1.56}$  &  $8.80_{-0.76}^{+0.88}$  &  3C186               &  1.063  &  1.826  &  $1.18_{-0.56}^{+0.86}$  &  ---                     \\
    \hline
  \end{tabular}
  }
\end{table*}
\renewcommand{\arraystretch}{1}

\label{lastpage}

\begin{thebibliography}{}

\bibitem[\protect\citeauthoryear{{Allen}, {Evrard}, \& {Mantz}}{{Allen}
  et~al.}{2011}]{Allen1103.4829}
{Allen} S.~W., {Evrard} A.~E.,  {Mantz} A.~B., 2011, \araa, in press,
  arXiv:1103.4829

\bibitem[\protect\citeauthoryear{{Allen} et~al.}{{Allen}
  et~al.}{2008}]{Allen0706.0033}
{Allen} S.~W., {Rapetti} D.~A., {Schmidt} R.~W., {Ebeling} H., {Morris} R.~G.,
  {Fabian} A.~C., 2008, \mnras, 383, 879

\bibitem[\protect\citeauthoryear{{Allen} et~al.}{{Allen}
  et~al.}{2004}]{Allen0405340}
{Allen} S.~W., {Schmidt} R.~W., {Ebeling} H., {Fabian} A.~C.,  {van Speybroeck}
  L., 2004, \mnras, 353, 457

\bibitem[\protect\citeauthoryear{{Allen}, {Schmidt}, \& {Fabian}}{{Allen}
  et~al.}{2001}]{Allen0110610}
{Allen} S.~W., {Schmidt} R.~W.,  {Fabian} A.~C., 2001, \mnras, 328, L37

\bibitem[\protect\citeauthoryear{{Ameglio} et~al.}{{Ameglio}
  et~al.}{2009}]{Ameglio0811.2199}
{Ameglio} S., {Borgani} S., {Pierpaoli} E., {Dolag} K., {Ettori} S.,  {Morandi}
  A., 2009, \mnras, 394, 479

\bibitem[\protect\citeauthoryear{{Arabadjis}, {Bautz}, \&
  {Arabadjis}}{{Arabadjis} et~al.}{2004}]{Arabadjis0305547}
{Arabadjis} J.~S., {Bautz} M.~W.,  {Arabadjis} G., 2004, \apj, 617, 303

\bibitem[\protect\citeauthoryear{{Arnaud}, {Pointecouteau}, \&
  {Pratt}}{{Arnaud} et~al.}{2005}]{Arnaud0502210}
{Arnaud} M., {Pointecouteau} E.,  {Pratt} G.~W., 2005, \aap, 441, 893

\bibitem[\protect\citeauthoryear{{Fabian} et~al.}{{Fabian}
  et~al.}{1981}]{Fabian1981ApJ...248...47F}
{Fabian} A.~C., {Hu} E.~M., {Cowie} L.~L.,  {Grindlay} J., 1981, \apj, 248, 47

\bibitem[\protect\citeauthoryear{{Fabian} et~al.}{{Fabian}
  et~al.}{2003}]{Fabian0306036}
{Fabian} A.~C., {Sanders} J.~S., {Allen} S.~W., {Crawford} C.~S., {Iwasawa} K.,
  {Johnstone} R.~M., {Schmidt} R.~W.,  {Taylor} G.~B., 2003, \mnras, 344, L43

\bibitem[\protect\citeauthoryear{{Finoguenov}, {Reiprich}, \&
  {B{\"o}hringer}}{{Finoguenov} et~al.}{2001}]{Finoguenov0010190}
{Finoguenov} A., {Reiprich} T.~H.,  {B{\"o}hringer} H., 2001, \aap, 368, 749

\bibitem[\protect\citeauthoryear{{Forman} et~al.}{{Forman}
  et~al.}{2005}]{Forman0312576}
{Forman} W. et~al., 2005, \apj, 635, 894

\bibitem[\protect\citeauthoryear{{Fukazawa}}{{Fukazawa}}{1997}]{Fukazawa1997Ph%
DT........24F}
{Fukazawa} Y., 1997, Ph.D. thesis, Univ.~Tokyo, (1997)

\bibitem[\protect\citeauthoryear{{Gelman} et~al.}{{Gelman}
  et~al.}{2004}]{Gelman2004BayesianDataAnalysis}
{Gelman} A., {Carlin} J.~B., {Stern} H.~S.,  {Rubin} D.~B., 2004, Bayesian Data
  Analysis.
\newblock Chapman \& Hall/CRC

\bibitem[\protect\citeauthoryear{{Horner}, {Mushotzky}, \& {Scharf}}{{Horner}
  et~al.}{1999}]{Horner9902151}
{Horner} D.~J., {Mushotzky} R.~F.,  {Scharf} C.~A., 1999, \apj, 520, 78

\bibitem[\protect\citeauthoryear{{Juett}, {Davis}, \& {Mushotzky}}{{Juett}
  et~al.}{2010}]{Juett0912.4078}
{Juett} A.~M., {Davis} D.~S.,  {Mushotzky} R., 2010, \apjl, 709, L103

\bibitem[\protect\citeauthoryear{{Kaiser}}{{Kaiser}}{1986}]{Kaiser1986MNRAS.22%
2..323K}
{Kaiser} N., 1986, \mnras, 222, 323

\bibitem[\protect\citeauthoryear{{Kelly}}{{Kelly}}{2007}]{Kelly0705.2774}
{Kelly} B.~C., 2007, \apj, 665, 1489

\bibitem[\protect\citeauthoryear{{Mahdavi} et~al.}{{Mahdavi}
  et~al.}{2008}]{Mahdavi0710.4132}
{Mahdavi} A., {Hoekstra} H., {Babul} A.,  {Henry} J.~P., 2008, \mnras, 384,
  1567

\bibitem[\protect\citeauthoryear{{Mantz} et~al.}{{Mantz}
  et~al.}{2008}]{Mantz0709.4294}
{Mantz} A., {Allen} S.~W., {Ebeling} H.,  {Rapetti} D., 2008, \mnras, 387, 1179

\bibitem[\protect\citeauthoryear{{Mantz} et~al.}{{Mantz}
  et~al.}{2010a}]{Mantz0909.3099}
{Mantz} A., {Allen} S.~W., {Ebeling} H., {Rapetti} D.,  {Drlica-Wagner} A.,
  2010a, MNRAS, 406, 1773

\bibitem[\protect\citeauthoryear{{Mantz} et~al.}{{Mantz}
  et~al.}{2010b}]{Mantz0909.3098}
{Mantz} A., {Allen} S.~W., {Rapetti} D.,  {Ebeling} H., 2010b, MNRAS, 406, 1759

\bibitem[\protect\citeauthoryear{{McNamara} \& {Nulsen}}{{McNamara} \&
  {Nulsen}}{2007}]{McNamara0709.2152}
{McNamara} B.~R.,  {Nulsen} P.~E.~J., 2007, \araa, 45, 117

\bibitem[\protect\citeauthoryear{{Morandi}, {Ettori}, \&
  {Moscardini}}{{Morandi} et~al.}{2007}]{Morandi0704.2678}
{Morandi} A., {Ettori} S.,  {Moscardini} L., 2007, \mnras, 379, 518

\bibitem[\protect\citeauthoryear{{Nagai}, {Vikhlinin}, \& {Kravtsov}}{{Nagai}
  et~al.}{2007}]{Nagai0609247}
{Nagai} D., {Vikhlinin} A.,  {Kravtsov} A.~V., 2007, \apj, 655, 98

\bibitem[\protect\citeauthoryear{{Navarro}, {Frenk}, \& {White}}{{Navarro}
  et~al.}{1997}]{Navarro9611107}
{Navarro} J.~F., {Frenk} C.~S.,  {White} S.~D.~M., 1997, \apj, 490, 493

\bibitem[\protect\citeauthoryear{{Nulsen}, {Powell}, \& {Vikhlinin}}{{Nulsen}
  et~al.}{2010}]{Nulsen1008.2393}
{Nulsen} P.~E.~J., {Powell} S.~L.,  {Vikhlinin} A., 2010, \apj, 722, 55

\bibitem[\protect\citeauthoryear{{Pfrommer} et~al.}{{Pfrommer}
  et~al.}{2007}]{Pfrommer0611037}
{Pfrommer} C., {En{\ss}lin} T.~A., {Springel} V., {Jubelgas} M.,  {Dolag} K.,
  2007, \mnras, 378, 385

\bibitem[\protect\citeauthoryear{{Pointecouteau}, {Arnaud}, \&
  {Pratt}}{{Pointecouteau} et~al.}{2005}]{Pointecouteau0501635}
{Pointecouteau} E., {Arnaud} M.,  {Pratt} G.~W., 2005, \aap, 435, 1

\bibitem[\protect\citeauthoryear{{Popesso} et~al.}{{Popesso}
  et~al.}{2005}]{Popesso0411536}
{Popesso} P., {Biviano} A., {B{\"o}hringer} H., {Romaniello} M.,  {Voges} W.,
  2005, \aap, 433, 431

\bibitem[\protect\citeauthoryear{{Rasia}, {Tormen}, \& {Moscardini}}{{Rasia}
  et~al.}{2004}]{Rasia0309405}
{Rasia} E., {Tormen} G.,  {Moscardini} L., 2004, \mnras, 351, 237

\bibitem[\protect\citeauthoryear{{Reiprich} \& {B{\"o}hringer}}{{Reiprich} \&
  {B{\"o}hringer}}{2002}]{Reiprich0111285}
{Reiprich} T.~H.,  {B{\"o}hringer} H., 2002, \apj, 567, 716

\bibitem[\protect\citeauthoryear{{Rozo} et~al.}{{Rozo}
  et~al.}{2010}]{Rozo0902.3702}
{Rozo} E. et~al., 2010, \apj, 708, 645

\bibitem[\protect\citeauthoryear{{Sarazin}}{{Sarazin}}{1988}]{Sarazin1988xrec.%
book.....S}
{Sarazin} C.~L., 1988, {X-ray emission from clusters of galaxies}

\bibitem[\protect\citeauthoryear{{Sasaki}}{{Sasaki}}{1996}]{Sasaki9611033}
{Sasaki} S., 1996, \pasj, 48, L119

\bibitem[\protect\citeauthoryear{{Schmidt} \& {Allen}}{{Schmidt} \&
  {Allen}}{2007}]{Schmidt0610038}
{Schmidt} R.~W.,  {Allen} S.~W., 2007, \mnras, 379, 209

\bibitem[\protect\citeauthoryear{{Schmidt}, {Allen}, \& {Fabian}}{{Schmidt}
  et~al.}{2001}]{Schmidt0107311}
{Schmidt} R.~W., {Allen} S.~W.,  {Fabian} A.~C., 2001, \mnras, 327, 1057

\bibitem[\protect\citeauthoryear{{Siemiginowska} et~al.}{{Siemiginowska}
  et~al.}{2010}]{Siemiginowska1008.1739}
{Siemiginowska} A., {Burke} D.~J., {Aldcroft} T.~L., {Worrall} D.~M., {Allen}
  S., {Bechtold} J., {Clarke} T.,  {Cheung} C.~C., 2010, \apj, 722, 102

\bibitem[\protect\citeauthoryear{{Simionescu} et~al.}{{Simionescu}
  et~al.}{2011}]{Simionescu1102.2429}
{Simionescu} A. et~al., 2011, \science, 331, 1576

\bibitem[\protect\citeauthoryear{{Sun} et~al.}{{Sun}
  et~al.}{2009}]{Sun0805.2320}
{Sun} M., {Voit} G.~M., {Donahue} M., {Jones} C., {Forman} W.,  {Vikhlinin} A.,
  2009, \apj, 693, 1142

\bibitem[\protect\citeauthoryear{{Vikhlinin} et~al.}{{Vikhlinin}
  et~al.}{2009a}]{Vikhlinin0805.2207}
{Vikhlinin} A. et~al., 2009a, \apj, 692, 1033

\bibitem[\protect\citeauthoryear{{Vikhlinin} et~al.}{{Vikhlinin}
  et~al.}{2006}]{Vikhlinin0507092}
{Vikhlinin} A., {Kravtsov} A., {Forman} W., {Jones} C., {Markevitch} M.,
  {Murray} S.~S.,  {Van Speybroeck} L., 2006, \apj, 640, 691

\bibitem[\protect\citeauthoryear{{Vikhlinin} et~al.}{{Vikhlinin}
  et~al.}{2009b}]{Vikhlinin0812.2720}
{Vikhlinin} A. et~al., 2009b, \apj, 692, 1060

\bibitem[\protect\citeauthoryear{{Voit}}{{Voit}}{2005}]{Voit0410173}
{Voit} G.~M., 2005, Reviews of Modern Physics, 77, 207

\bibitem[\protect\citeauthoryear{{White}, {Jones}, \& {Forman}}{{White}
  et~al.}{1997}]{White1997MNRAS.292..419W}
{White} D.~A., {Jones} C.,  {Forman} W., 1997, \mnras, 292, 419

\bibitem[\protect\citeauthoryear{{Wu}, {Rozo}, \& {Wechsler}}{{Wu}
  et~al.}{2010}]{Wu0907.2690}
{Wu} H., {Rozo} E.,  {Wechsler} R.~H., 2010, \apj, 713, 1207

\end{thebibliography}
\end{document}